\documentstyle[aps]{revtex}
\begin{document}
\draft
\title{Quantum information is incompressible without errors}
\author{Masato Koashi and Nobuyuki Imoto}
\address{CREST Research Team for Interacting Carrier Electronics,
School of
Advanced Sciences, \\
~The Graduate University for Advanced Studies (SOKEN),
Hayama, Kanagawa, 240-0193, Japan}
\maketitle
\begin{abstract}
A classical random variable can be faithfully 
compressed into a sequence of bits with its expected length 
lies within one bit of Shannon entropy. We generalize 
this variable-length and faithful scenario to the general 
quantum source producing mixed states $\rho_i$ with 
probability $p_i$. In contrast to the classical case, 
the optimal compression rate in the limit of large block length 
differs from the one in the fixed-length and 
asymptotically faithful scenario. The amount
 of this gap is interpreted as the genuinely quantum 
part being incompressible in the former scenario.

\end{abstract}
 

One of the fundamental questions in the information theory is 
about the data compression, namely, what is 
the shortest description of a data. This question is important 
not only for quantification of the amount of information,
but also  
for understanding how well we can manipulate information
stored in physical systems, which is a central topic in the 
field of quantum information.
In the case of compressing 
a classical random variable $X$ which takes 
letter $x$ with probability 
$p(x)$, the answer is given by the quantity called the 
Shannon entropy, which is defined as 
$H(X)=-\sum_x p(x)\log_2 p(x)$ when measured in bits \cite{shannon48}.
More precisely,
a compression scheme is regarded as an assignment of each letter $x$ 
with a codeword of length $l(x)$ bits, which is not fixed and dependent on 
$x$. 
 Under the requirement that 
the codewords are uniquely decodable (faithful)
 even when they are concatenated
to describe a sequence of letters, the minimum $L_{\rm min}$ of 
the expected length $L\equiv \sum_x p(x)l(x)$ satisfies \cite{cover}
$H(X)\le L_{\rm min} \le H(X)+1$.
When a sequence of $n$ letters independently drawn according to 
$p(x)$ is collectively compressed, the minimum 
expected length lies between $nH(X)$ and $nH(X)+1$ 
because of the additivity of the entropy function.
Hence the optimum compression rate $R^{(n)}_{\rm opt}$
for block length $n$,
defined as the minimum 
expected length per letter, satisfies
\begin{equation}
H(X)\le R^{(n)}_{\rm opt} \le H(X)+1/n,
\label{class_n}
\end{equation} 
and thus
$R^{(\infty)}_{\rm opt}\equiv 
\lim_{n\rightarrow \infty}R^{(n)}_{\rm opt} =H(X)$. 

In addition to this `variable-length and faithful' (VLF) scenario,
we encounter the Shannon entropy also in a slightly different 
scenario of compression, in which 
all codewords have a common fixed length, 
and small errors 
in decoding are allowed as long as they vanish in 
 the limit of large block-length. 
In this 
`fixed-length and asymptotically faithful' (FLAF)
scenario, we consider a sequence of compression schemes labeled by 
the block length $n$, which have code length $L^{(n)}$
and error probability $p^{(n)}_{\rm err}$. 
The sequence is asymptotically faithful when 
$\lim_{n\rightarrow \infty}p^{(n)}_{\rm err}= 0$. The
asymptotic rate of compression for the sequence is characterized by
$R\equiv \overline{\lim}_{n\rightarrow \infty}L^{(n)}/n$.
The optimal compression rate $R^{\rm AF}_{\rm opt}$
in this scenario is defined as the infimum of $R$ among all
 asymptotically faithful sequences. Requirement for 
this scenario is weaker than the VLF scenario in spite of 
the constraint on the codeword length. This is seen by 
composing a FLAF sequence with asymptotic rate $R^{(n)}_{\rm opt}+\delta$
by just repeating $m$ times a VLF scheme
with rate $R^{(n)}_{\rm opt}$ and treat the concatenated codewords
longer than $mn(R^{(n)}_{\rm opt}+\delta)$ as errors. 
With $n$ fixed and in the limit of $m\rightarrow\infty$,
this sequence is asymptotically faithful for any $\delta>0$
due to the law of large numbers. Hence we have 
\begin{equation}
R^{(n)}_{\rm opt}\ge R^{\rm AF}_{\rm opt},
\label{F_ge_AF}
\end{equation}
and $R^{(\infty)}_{\rm opt}\ge R^{\rm AF}_{\rm opt}$.
In the case of compressing 
a classical random variable $X$, 
it is known that $R^{\rm AF}_{\rm opt}$ is 
also equal to the Shannon entropy, namely,
$R^{(\infty)}_{\rm opt}= R^{\rm AF}_{\rm opt}=H(X)$ \cite{shannon48,cover}.

In the quantum case of compressing an ensemble 
${\cal E}=\{p_i, \hat\rho_i\}$, we consider 
a source that emits a quantum system in state (letter) $\hat\rho_i$ with probability
$p_i$, and $n$ systems drawn from this source are compressed into 
qubits. The discussions were first focused on the FLAF 
scenario, and the result when 
$\hat\rho_i$ are all pure states \cite{schumacher95,jozsa94}
is $R^{\rm AF}_{\rm opt}=S(\hat\rho)$, 
where $S(\hat\rho)\equiv -{\rm Tr}[\hat\rho\log_2\hat\rho]$ is the von Neumann 
entropy of the average state $\hat\rho\equiv \sum_i p_i \hat\rho_i$. 
When $\{\hat\rho_i\}$ includes mixed states, $R^{\rm AF}_{\rm opt}$
is still given by the von Neumann entropy after removing hidden
redundancy \cite{compressibility}.
 This striking similarity to the classical case implies that
there is no big difference in the compressibility between classical and 
quantum information, at least in the FLAF scenario. 
Recently, the investigations on different scenarios 
\cite{horodecki98,horodecki99,barnum00,dur01,kramer01,hayashi02}
have been started to
 reveal how the optimal rates vary depending on small differences 
in the constraints. A notable example is the information defect, the
difference between the rate $R^{\rm AF}_{\rm opt}$ defined above 
(which is referred to as the blind scenario in this context)
 and the rate in 
an easier scenario (the visible scenario) 
in which the identity of the index $i$ is available 
in the compression stage. The information defect has turned out to be
nonzero \cite{compressibility,kramer01}
even in the classical cases where all $\{\hat\rho_i\}$ commute, 
implying that the origin of this gap does not necessarily lie in the 
nature of quantum information.

In this Letter, we discuss the VLF scenario for general quantum 
ensemble ${\cal E}=\{p_i, \hat\rho_i\}$, and derive inequalities 
corresponding to Eq.~(\ref{class_n}), which identify
$R^{(\infty)}_{\rm opt}$. We show that the gap 
$\Delta_{\rm F-AF}\equiv R^{(\infty)}_{\rm opt}-R^{\rm AF}_{\rm opt}$
is generally nonzero, and nonzero only if $\{\hat\rho_i\}$ do not commute.
When we further separate the information in ${\cal E}$ into 
 the classical part
 and the quantum part, the origin of the gap becomes transparent, namely,
the genuinely quantum part of the information turns out to be 
incompressible in the VLF scenario.

The keystone in our derivation of $R^{(\infty)}_{\rm opt}$ is the
observation that the length of the qubits used in a single round of 
the compression operation can be regarded as an outcome of a measurement 
on the system to be compressed. This is justified in an operational sense 
as follows. Suppose that initially there is a resource of $N$ qubits,
and an input state from the source ${\cal E}=\{p_i, \hat\rho_i\} (p_i>0)$  
acting on Hilbert 
space ${\cal H}'_A$
 are compressed into a number of qubits via a VLF scheme.
The notion that ``$L$ qubits has been used to compress the input state'' 
means that there should remain $N-L$ qubits in the resource which can 
be utilized in other {\it independent} tasks.
 This means that after the state is decompressed back in ${\cal H}'_{\rm A}$,
the value $L$ can be determined without accessing ${\cal H}'_{\rm A}$. In other 
words, if we generally write the whole quantum operation of the compression and 
decompression as a unitary operation $\hat{U}$ on the combined system of 
${\cal H}'_{\rm A}$ and an auxiliary system ${\cal H}_{\rm E}$ initially 
prepared in a standard pure state $\hat\Sigma_{\rm E}$, 
we should be able to define an observable $\hat{L}$ acting on ${\cal H}_{\rm E}$,
which corresponds to the length of qubits used to compress.
The expected length $\langle \hat{L} \rangle$ of the compression scheme $\hat{U}$
 applied to the source 
${\cal E}$ is then written as
\begin{equation}
\langle \hat{L} \rangle=
{\rm Tr}_{\rm E}\{
\hat{L}{\rm Tr}_{\rm A}
[\hat{U}(\hat\rho\otimes\hat\Sigma_{\rm E})\hat{U}^\dagger]\},
\label{ave_L}
\end{equation}
where $\hat\rho\equiv \sum_i p_i \hat\rho_i$.
At the same time, since the compression scheme is faithful (no errors),
$\hat{U}$ obeys
\begin{equation}
{\rm Tr}_{\rm E}
[\hat{U}(\hat\rho_i\otimes\hat\Sigma_{\rm E})\hat{U}^\dagger]\}=\hat\rho_i.
\label{no_disturb}
\end{equation}
The length of qubits is thus regarded as the outcome of a generalized measurement on 
${\cal H}'_{\rm A}$ that introduces no disturbance on the initial states 
$\{\hat\rho_i\}$.

The property of the operations that preserves the initial states 
$\{\hat\rho_i\}$ was 
analyzed in detail recently\cite{what}, and 
it was shown that,
given $\{\hat\rho_i\}$,
we can find a unique decomposition of ${\cal H}_{\rm A}$,
the support of $\hat\rho$,
  written as
\begin{equation}
{\cal H}_{\rm A}=\bigoplus_l
{\cal H}^{(l)}_{\rm J} \otimes {\cal H}^{(l)}_{\rm K}.
\label{hdec}
\end{equation}
Under this decomposition, $\hat\rho_i$
is written as
\begin{equation}
\hat\rho_i=\bigoplus_l p^{(i,l)} 
\hat\rho^{(i,l)}_{\rm J}\otimes \hat\rho^{(l)}_{\rm K},
\label{rhodec}
\end{equation}
where $\hat\rho^{(i,l)}_{\rm J}$ and $\hat\rho^{(l)}_{\rm K}$ are normalized density 
operators acting on ${\cal H}^{(l)}_{\rm J}$ and ${\cal H}^{(l)}_{\rm K}$, respectively, 
and
$p^{(i,l)}$ is the probability for the state to be  in the subspace
${\cal H}^{(l)}_{\rm J}\otimes{\cal H}^{(l)}_{\rm K}$.
 $\hat\rho^{(l)}_{\rm K}$ is independent  of $i$, and
$\{\hat\rho^{(1,l)}_{\rm J},\hat\rho^{(2,l)}_{\rm J},\ldots\}$ cannot be
expressed in a simultaneously block-diagonalized form. 
Any $\hat{U}$ that satisfies Eq.~(\ref{no_disturb})
is then expressed in the following form
\begin{equation}
\hat{U}(\hat{1}_{\rm A}\otimes \hat\Sigma_{\rm E})
=\bigoplus_l\hat{1}^{(l)}_{\rm J}\otimes \hat{U}^{(l)}_{\rm
KE}(\hat{1}^{(l)}_{\rm K}\otimes \hat\Sigma_{\rm E}), 
\label{main1}
\end{equation}
where $\hat{U}^{(l)}_{\rm KE}$ are unitary operators acting on the combined
space ${\cal H}^{(l)}_{\rm K}\otimes {\cal H}_{\rm E}$,
satisfying 
${\rm Tr}_{\rm E}
[\hat{U}^{(l)}_{\rm KE}(\hat\rho^{(l)}_{\rm K}\otimes\hat\Sigma_{\rm E})
\hat{U}^{(l)\dagger}_{\rm KE}]\}=\hat\rho^{(l)}_{\rm K}$.
An explicit procedure to obtain this particular decomposition is also 
given in \cite{what}.

The total density operator $\hat\rho\equiv \sum_i p_i \hat\rho_i$ for the
ensemble ${\cal E}$ is also decomposed as 
\begin{equation}
\hat\rho=\bigoplus_l p^{(l)} \hat\rho^{(l)}_{\rm J}\otimes \hat\rho^{(l)}_{\rm K},
\label{totalrhodec}
\end{equation}
where $p^{(l)}\equiv\sum_i p_i p^{(i,l)}$ and 
$\hat\rho^{(l)}_{\rm J}\equiv (\sum_i p_i p^{(i,l)}
\hat\rho^{(i,l)}_{\rm J})/p^{(l)}$.
Now, let us take a basis $\{|a_j\rangle^{(l)}_{\rm J}\}$ 
$(j=1,\ldots, {\rm dim}\; {\cal H}^{(l)}_{\rm J})$ for each 
${\cal H}^{(l)}_{\rm J}$, and 
consider another source ${\cal E}'=\{q_\lambda, \hat\sigma_\lambda\}$
with double index $\lambda\equiv (l,j)$,
which is defined through the decomposition (\ref{totalrhodec}) such that 
$q_\lambda=({\rm dim}\; {\cal H}^{(l)}_{\rm J})^{-1}p^{(l)}$ and 
\begin{equation}
\hat\sigma_\lambda=  
|a_j\rangle^{(l)}_{\rm J}{}^{(l)}_{\rm J}\langle a_j|\otimes \hat\rho^{(l)}_{\rm K}.
\label{sigma}
\end{equation}
The total density operator $\hat\sigma\equiv \sum_\lambda q_\lambda \hat\sigma_\lambda$ 
for this source is
\begin{equation}
\hat\sigma=\bigoplus_l p^{(l)} ({\rm dim}\; {\cal H}^{(l)}_{\rm J})^{-1}
\hat{1}^{(l)}_{\rm J}\otimes \hat\rho^{(l)}_{\rm K}.
\label{totalsigmadec}
\end{equation}
Suppose that this source is used as an input to the same compression scheme.
The form (\ref{main1}) assures that the operation $\hat{U}$ also 
preserves $\hat\sigma_\lambda$, namely, the scheme $\hat{U}$ works as 
a VLF scheme for ${\cal E}'$.
 We further note that, since the difference 
between $\hat\rho$ and $\hat\sigma$ lies in the internal states of
${\cal H}^{(l)}_{\rm J}$, and the form (\ref{main1}) ensures that the 
state of the auxiliary system (${\cal H}_{\rm E}$) after the operation 
$\hat{U}$ is insensitive to them, we obtain
${\rm Tr}_{\rm A}
[\hat{U}(\hat\rho\otimes\hat\Sigma_{\rm E})\hat{U}^\dagger]=
{\rm Tr}_{\rm A}
[\hat{U}(\hat\sigma\otimes\hat\Sigma_{\rm E})\hat{U}^\dagger]$.
 Then, according to Eq.~(\ref{ave_L}),
the same expected length $\langle \hat{L} \rangle$ should be observed 
even when the source is replaced with ${\cal E}'$. Hence we have
\begin{equation}
L_{\rm min}({\cal E})\ge L_{\rm min}({\cal E}').
\label{E_ge_Ep}
\end{equation}

In order to find a lower bound of $L_{\rm min}({\cal E}')$,
we use Eq.~(\ref{F_ge_AF}), which also holds for general 
quantum ensembles. 
Since the letter states $\hat\sigma_\lambda$
of the ensemble ${\cal E}'$ are all orthogonal,
the optimum rate $R^{\rm AF}_{\rm opt}({\cal E}')$  should 
be equal to that of an 
orthogonal pure-state ensemble $\{q_\lambda, |\lambda\rangle\}$,
and hence $R^{\rm AF}_{\rm opt}({\cal E}')=-\sum_\lambda q_\lambda
\log_2 q_\lambda$.
Using this with $q_\lambda=({\rm dim}\; {\cal H}^{(l)}_{\rm J})^{-1}p^{(l)}$,
Eq.~(\ref{F_ge_AF}) with $n=1$, and 
$L_{\rm min}({\cal E}')=R^{(1)}_{\rm opt}({\cal E}')$ (by definition),
we have
\begin{equation}
L_{\rm min}({\cal E}')\ge R^{\rm AF}_{\rm opt}({\cal E}')=
I_{\rm C}({\cal E})+D_{\rm NC}({\cal E}),
\label{Ep_ge_AFp}
\end{equation}
where $I_{\rm C}({\cal E})\equiv -\sum_l p^{(l)}
\log_2 p^{(l)}$ and $D_{\rm NC}({\cal E})\equiv
\sum_l p^{(l)} \log_2 {\rm dim}\;{\cal H}^{(l)}_{\rm J}$.

For an upper bound for $L_{\rm min}({\cal E})$, we consider a
specific example of 
the VLF scheme as follows. The compressor first makes a projection 
measurement to determine the index $l$, 
which appears in the decomposition (\ref{hdec}). 
The projected state should be supported
by subspace ${\cal H}^{(l)}_{\rm J}\otimes {\cal H}^{(l)}_{\rm K}$.
Since the outcome $l$ 
occurs with probability $p^{(l)}$,
there exists an instantaneous code for encoding $l$ with an expected length 
not larger than $H(\{p^{(l)}\})+1$ bits \cite{cover}. The compressor
write down this classical codeword onto qubits using the standard basis
$\{|0\rangle,|1\rangle\}$. It then discards ${\cal H}^{(l)}_{\rm K}$,
and transfer the state of ${\cal H}^{(l)}_{\rm J}$ (which should be 
$\hat\rho^{(i,l)}_{\rm J}$) into 
$\lceil\log_2 {\rm dim}\;{\cal H}^{(l)}_{\rm J} \rceil$ qubits
and concatenate them after the qubits holding the classical 
codeword for $l$. The expected total length of the qubits is then 
not larger than $I_{\rm C}({\cal E})+1+D_{\rm NC}({\cal E})+1$.
The decompression is done by measuring qubits in the standard basis
one by one, until the end of the codeword is reached. Note that the decompressor
knows this end point since the code for $l$ 
is instantaneous. Learning $l$, it then transfers the contents of the next 
$\lceil\log_2 {\rm dim}\;{\cal H}^{(l)}_{\rm J} \rceil$ qubits 
into ${\cal H}^{(l)}_{\rm J}$, and prepare ${\cal H}^{(l)}_{\rm K}$
 in the known state $\hat\rho^{(l)}_{\rm K}$. When the input to 
the compressor was $\hat\rho_i$, the conditional probability
of $l$ is $p^{(i,l)}$. Hence  
the whole process faithfully reproduces $\hat\rho_i$ [see Eq.~(\ref{rhodec})]. We thus 
obtain an upper bound $I_{\rm C}({\cal E})+D_{\rm NC}({\cal E})+2$
for $L_{\rm min}({\cal E})$. Together with Eqs.~(\ref{E_ge_Ep}) and
(\ref{Ep_ge_AFp}), we have
\begin{equation}
I_{\rm C}({\cal E})+D_{\rm NC}({\cal E}) \le 
L_{\rm min}({\cal E})\le I_{\rm C}({\cal E})+D_{\rm NC}({\cal E})+2.
\label{result1}
\end{equation}

In order to extend this result to the case of $n$-block coding,
we need to know the behavior of the functions 
$I_{\rm C}({\cal E})$ and $D_{\rm NC}({\cal E})$
for the concatenation of independently drawn letters (states).
Suppose that ${\cal H}_{\rm A}$ is prepared by a source 
${\cal E}_{\rm A}=\{p^{\rm A}_i,\hat\rho^{\rm A}_i\}$,
and let ${\cal H}_{\rm A}=\bigoplus_l
{\cal H}^{(l)}_{\rm AJ} \otimes {\cal H}^{(l)}_{\rm AK}$ be the 
decomposition determined by ${\cal E}_{\rm A}$.
Suppose further that another system ${\cal H}_{\rm B}$ 
is independently prepared by a source 
${\cal E}_{\rm B}=\{p^{\rm B}_j,\hat\rho^{\rm B}_j\}$,
and let ${\cal H}_{\rm B}=\bigoplus_m
{\cal H}^{(m)}_{\rm BJ} \otimes {\cal H}^{(m)}_{\rm BK}$ be the 
decomposition determined by ${\cal E}_{\rm B}$.
This preparation can be also considered as the combined system 
${\cal H}_{\rm AB}\equiv{\cal H}_{\rm A}\otimes {\cal H}_{\rm B}$ being prepared by 
${\cal E}_{\rm AB}=\{p^{\rm A}_ip^{\rm B}_j,
\hat\rho^{\rm A}_i\hat\rho^{\rm B}_j\}$. It was shown \cite{what}
that the decomposition determined by ${\cal E}_{\rm AB}$ is
simply given by the direct product 
${\cal H}_{\rm AB}=\bigoplus_{\{l,m\}}
({\cal H}^{(l)}_{\rm AJ}\otimes{\cal H}^{(m)}_{\rm BJ}) 
\otimes ({\cal H}^{(l)}_{\rm AK}\otimes{\cal H}^{(m)}_{\rm BK})$.
This implies that $I_{\rm C}({\cal E})$ and $D_{\rm NC}({\cal E})$ are 
additive, namely, $I_{\rm C}({\cal E}_{\rm AB})=I_{\rm C}({\cal E}_{\rm A})+
I_{\rm C}({\cal E}_{\rm B})$ and $D_{\rm NC}({\cal E}_{\rm AB})=D_{\rm NC}({\cal E}_{\rm A})+
D_{\rm NC}({\cal E}_{\rm B})$. Just as in the classical case, 
this additivity leads to our main results,
\begin{equation}
I_{\rm C}({\cal E})+D_{\rm NC}({\cal E}) \le 
R^{(n)}_{\rm opt}({\cal E})\le I_{\rm C}({\cal E})+D_{\rm NC}({\cal E})+2/n,
\label{result2}
\end{equation}
and 
\begin{equation}
R^{(\infty)}_{\rm opt}({\cal E})=I_{\rm C}({\cal E})+D_{\rm NC}({\cal E}).
\label{result3}
\end{equation}

This optimal compression rate $R^{(\infty)}_{\rm opt}({\cal E})$
 for the VLF scenario is generally larger than the one in the FLAF scenario,
which was derived in \cite{compressibility} for general mixed-state cases as 
\begin{equation}
R^{\rm AF}_{\rm opt}({\cal E})=
I_{\rm C}({\cal E})+I_{\rm NC}({\cal E}),
\label{resultAF}
\end{equation}
where $I_{\rm NC}({\cal E})\equiv
\sum_l p^{(l)} S(\rho^{(l)}_{\rm J})$, and
$I_{\rm C}({\cal E})= -\sum_l p^{(l)}
\log_2 p^{(l)}$ is the same function as defined above.
The difference
$\Delta_{\rm F-AF}\equiv R^{(\infty)}_{\rm opt}-R^{\rm AF}_{\rm opt}$ is
\begin{equation}
\Delta_{\rm F-AF}=D_{\rm NC}({\cal E})-I_{\rm NC}({\cal E})
=\sum_l p^{(l)} [\log_2 {\rm dim}\;{\cal H}^{(l)}_{\rm J}-S(\hat\rho^{(l)}_{\rm J})],
\label{d_F_AF}
\end{equation}
which is nonzero when there exists $l$ such that 
$\hat\rho^{(l)}_{\rm J}\neq ({\rm dim}\;{\cal H}^{(l)}_{\rm J})^{-1}
\hat{1}^{(l)}_{\rm J}$. The gap $\Delta_{\rm F-AF}$ is zero 
when $\hat\rho^{(l)}_{\rm J}=
({\rm dim}\;{\cal H}^{(l)}_{\rm J})^{-1}\hat{1}^{(l)}_{\rm J}$ 
for all $l$. In particular, for the classical 
cases in which all $\hat\rho_i$ commutes, 
$({\rm dim}\;{\cal H}^{(l)}_{\rm J})=1$ for all $l$ \cite{what}
and $\Delta_{\rm F-AF}=0$.

The gap $\Delta_{\rm F-AF}$ stems from the internal state 
of each ${\cal H}^{(l)}_{\rm J}$, which is regarded as 
the genuinely quantum (nonclassical) part of the ensemble ${\cal E}$ 
in the following sense: (i) This part is inaccessible without 
introducing disturbance [Eq.~(\ref{main1})] and (ii) 
this part can be compressed only into qubits, not into classical bits
\cite{jozsa,cost}. Then, the present results are summarized in a
simple statement, `the genuinely quantum part of the information is 
incompressible if no errors are allowed.' In contrast, the classical 
part of the information, which is represented by $\{p^{(i,l)}\}$,
can be compressed into $I_{\rm C}$ bits in either one of the scenarios. 
The difference between the  classical and 
the quantum part in the nature of compressibility may be understood as
follows. The classical information can be effectively compressed by 
changing the description length adaptively depending on the input
states, namely, using shorter descriptions for frequent inputs and 
longer descriptions for rare inputs.
The quantum information, if it is to be compressed faithfully,
does not allow such adaptation because learning the input state 
inevitably introduces irreversible disturbance.

It may be quite instructive to contrast the derived gap $\Delta_{\rm F-AF}$
with the information defect, the gap between the blind and the visible
scenario. In the visible scenario, the classical index $i$ is given to the
compressor, rather than the state $\rho_i$ to be reproduced in the decompression. 
The optimal compression rate $I_{\rm eff}$ for 
 asymptotically faithful schemes in the visible scenario is called the 
effective information. By definition, $R^{\rm AF}_{\rm opt} \ge I_{\rm eff}$,
and the gap $\Delta_{\rm b-v}\equiv R^{\rm AF}_{\rm opt} - I_{\rm eff}\ge 0$
is called the information defect. 
While the explicit form of the 
effective information is still open, it is bounded  
 from below by
the Levitin-Holevo function \cite{holevo73} 
$I_{\rm LH}\equiv S(\hat\rho)-\sum_i p_i S(\hat\rho_i)$,
namely, $I_{\rm eff}\ge I_{\rm LH}$ \cite{horodecki98}.
An upper bound for $\Delta_{\rm b-v}$ is hence 
$R^{\rm AF}_{\rm opt} - I_{\rm LH}=I_{\rm C}+I_{\rm NC}-I_{\rm LH}$,
and using Eqs.~(\ref{rhodec}) and (\ref{totalrhodec}), we obtain 
\begin{equation}
 \Delta_{\rm b-v}\le \sum_i p_i S(\hat\rho_i)-\sum_l p^{(l)}S(\hat\rho^{(l)}_{\rm K})
=\sum_i p_i S(\hat\rho_i^{\rm R}),
\label{b_f_le}
\end{equation}
where the state $\hat\rho_i^{\rm R}\equiv \bigoplus_l p^{(i,l)} 
\hat\rho^{(i,l)}_{\rm J}$ is the one obtained 
from $\hat\rho_i$ by removing the redundant part $\hat\rho^{(l)}_{\rm K}$
[see Eq.(\ref{rhodec})]. Let us call an ensemble ${\cal E}$ `pure' when 
all the states $\hat\rho_i^{\rm R}$ are pure states, and call it 
`mixed' otherwise. Then, Eq.~(\ref{b_f_le}) implies that 
$\Delta_{\rm b-v}=0$ for pure ensembles.
It should be noted that the pure/mixed classification is independent 
of the one by classical (all $\rho_i$ commutes) and quantum.
It has been shown \cite{compressibility,kramer01} that there exist classical ensembles with
$\Delta_{\rm b-v}>0$. The properties of the two gaps 
$\Delta_{\rm F-AF}$ and $\Delta_{\rm b-v}$ are summarized 
in Table \ref{tab:1}.
When the letter states are distinct, namely, 
$\{\hat\rho_i\}$ are all orthogonal, the optimal compression rates 
for the three scenarios are equal. When 
the letter states overlap with each other and 
are not completely distinguishable, 
there can be two types of overlaps. One is the cases when 
each letter state is noisy. In such cases, 
whether or not the identity of each letter is given as side information 
affect the compressibility.
The other way of overlapping is quantum one, 
where the letter states are pure and suffers no noises, 
but are not completely distinguishable from each other due to 
nonorthogonality. In such cases, the 
compressibility depends on whether small 
errors (even an asymptotically vanishing one) are allowed or not.

This work was supported by a Grant-in-Aid for Encouragement of Young
Scientists (Grant No.~12740243) and a Grant-in-Aid for Scientific 
Research (B) (Grant No.~12440111)
by Japan Society for the Promotion of
Science.

\begin{table}
\begin{center}
\begin{tabular}{ccc} 
Information & \begin{tabular}{c} classical \\ ($[\hat\rho_i,\hat\rho_j]=0$)
\end{tabular}&
quantum 
\\\hline
\begin{tabular}{c} pure \\ ($S(\hat\rho^{\rm R}_i)=0$) \end{tabular}&
$R^{(\infty)}_{\rm opt}=R^{\rm AF}_{\rm opt}= I_{\rm eff}$&
$R^{(\infty)}_{\rm opt}\ge R^{\rm AF}_{\rm opt}= I_{\rm eff}$\\
mixed & $R^{(\infty)}_{\rm opt}=R^{\rm AF}_{\rm opt}\ge I_{\rm eff}$&
$R^{(\infty)}_{\rm opt}\ge R^{\rm AF}_{\rm opt}\ge I_{\rm eff}$\\
\end{tabular}
\end{center}
\caption{Nature of information and 
gaps between optimal compression rates for various scenarios
--- the variable-length, faithful and blind scenario 
($R^{(\infty)}_{\rm opt}$),
the fixed-length, asymptotically faithful and blind scenario
($R^{\rm AF}_{\rm opt}$), and the 
fixed-length, asymptotically faithful and visible scenario
($I_{\rm eff}$).
\label{tab:1}}
\end{table}

\end{document}